# Approximate Reservoir Computing with a Semiconductor Laser for Reducing Energy Consumption

Tatsuki Ito, Kazutaka Kanno, Satoshi Kawakami, *Member, IEEE,* and Atsushi Uchida, *Senior Member, IEEE*

***Abstract*— Photonic reservoir computing is a promising physical machine-learning approach for time-series prediction. Quantization of the reservoir response signal is an important consideration in the implementation of photonic reservoir computing, and the number of quantization bits and sampling frequency need to be optimized to achieve high prediction performance and low energy consumption. However, few studies have investigated the effects of the number of quantization bits and sampling frequency on both prediction performance and energy consumption. In this study, we introduce the concept of approximate reservoir computing with a semiconductor laser by quantizing the reservoir node states and output weights. We numerically evaluate the performance of a chaotic time-series prediction task and estimate the energy consumption per sample using a component-level model. By optimizing the number of quantization bits, the sampling frequency, and the injection current of the semiconductor laser, we achieve a significant reduction in the estimated energy consumption per sample while maintaining prediction performance. We also obtain model-based estimates of the throughput, energy efficiency, and energy per operation of the proposed reservoir-computing system.**

## I. INTRODUCTION

PHOTONIC artificial intelligence has attracted increasing interest as a high-speed and energy-efficient machine-learning paradigm, known as a photonic accelerator [1]. Photonic neural networks are promising approaches for photonic accelerators [2]. Recently, recurrent neural networks have been used for predicting time-series data, because they have a structure that retains past information (short-term memory) in the hidden layers. However, they have the disadvantage of high computational cost due to the large number of training parameters and the difficulty of parallelization.

To solve this issue, reservoir computing has attracted attention as a machine-learning scheme with low computational cost [3,4]. In reservoir computing, the weights of the input and reservoir parts are fixed, and only the weights of the output part are optimized by learning. A single nonlinear element with time-delay feedback can be utilized as the reservoir part, and various photonic implementations have been reported [5]. In particular, implementations using semiconductor lasers are capable of high-speed information processing on the order of GHz [6].

The concept of approximate computing has been proposed [7], where the effective number of bits for calculation is limited to efficiently utilize computational resources and reduce energy consumption. Although conventional computing architectures are designed for high-precision computation, strict accuracy may not be required in some applications. For deep learning, it has been reported that high accuracy is not always needed for intermediate-layer computations [8].

Physical systems such as photonic reservoir computing using a semiconductor laser require quantization of the response signal (node states) through Analog-to-Digital conversion. Therefore, the quantization bits can be adjusted by applying the concept of approximate computing. However, few studies have been reported to investigate the relationship between the number of quantization bits and prediction performance in photonic reservoir computing [9]. Moreover, there has been limited knowledge of the quantitative estimation of energy consumption in photonic reservoir computing [10,11].

In this study, we introduce a concept of approximate reservoir computing with a semiconductor laser. We change the number of quantization bits of the response signal from the reservoir and evaluate the performance of a chaotic time-series prediction task. We also evaluate the energy consumption per sample of the photonic approximate reservoir computing. We optimize the parameter values of the number of quantization bits, the sampling frequency of the Analog-to-Digital Converter (ADC), and the injection current of the semiconductor laser to reduce the energy consumption. We also evaluate the throughput, energy efficiency, and energy per operation in the proposed reservoir-computing system quantitatively.

This work was supported in part by JSPS KAKENHI (Grant Numbers JP22H05195, JP25H01129) and JST, CREST (Grant Number JPMJCR24R2).
*(Corresponding author: Atsushi Uchida).*

Tatsuki Ito, Kazutaka Kanno, and Atsushi Uchida are with the Department of Information and Computer Sciences, Saitama University, 255 Shimo-okubo, Sakura-ku, Saitama City, Saitama 338-8570, Japan (e-mails: jam079@docomo.ne.jp, kkanno@mail.saitama-u.ac.jp, auchida@mail.saitama-u.ac.jp).

Satoshi Kawakami is with the Faculty of Information Science and Electrical Engineering, Kyushu University, 744 Motooka, Nishi-ku, Fukuoka 819-0395, Japan (e-mail: kawakami@ed.kyushu-u.ac.jp).

## II. Scheme for Approximate Reservoir Computing

We use photonic reservoir computing using a semiconductor laser with feedback light, which is known as delay-based reservoir computing [5]. The scheme for delay-based reservoir computing consists of three parts: an input part, a reservoir part, and an output part. The roles of each part are described below.

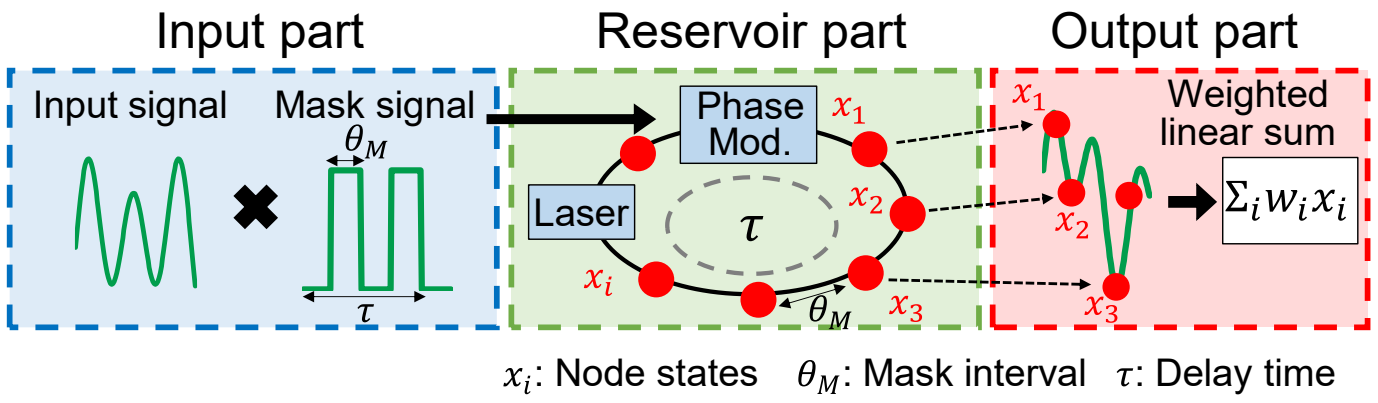

**Fig. 1.** Schematic for reservoir computing.

### *A. Input part*

The input part preprocesses the modulation signal by multiplying the input signal $u(t)$ by the mask signal $m(t)$ with a scaling factor $\gamma$. The modulation signal $s(t)$ is expressed by the following equation.

$$s(t) = \gamma u(t) m(t), \tag{1}$$

The discrete input data $u(n)$ is converted to the continuous input signal $u'(n)$ stretched by the feedback delay time $\tau$.

$$u(t) = u'(n), \qquad \big((n-1)\tau \le t < n\tau\big), \tag{2}$$

where $n$ is the discrete-time index of the input data and $t$ is the continuous-time variable.

The mask signal $m(t)$ is generated by creating a four-valued mask signal $m'(n) \in \{\pm 0.3, \pm 1.0\}$ for the number of nodes, which is then stretched by the mask interval $\theta_\mathrm{M}$ to generate a continuous-time mask signal $m(t)$.

$$m(t) = m'(i), \qquad \big((i-1)\theta_\mathrm{M} \le t < i\theta_\mathrm{M}\big), \tag{3}$$

$$m(t+\tau) = m(t), \tag{4}$$

where $i$ is the integer index of the mask interval, ranging from 1 to $I_{max}$. Here, $I_{max} = \lfloor \tau/\theta_M \rfloor$ denotes the maximum integer of $\tau/\theta_M$.

### *B. Reservoir part of semiconductor laser with feedback light*

The reservoir part consists of a single semiconductor laser with time-delay feedback. The following equations express the Lang-Kobayashi equations with time-delay feedback [12-14].

$$\frac{dE(t)}{dt} = \frac{1+i\alpha}{2}\left[\frac{G_N(N(t)-N_0)}{1+\varepsilon|E(t)|^2} - \frac{1}{\tau_p}\right]E(t) + \kappa E^{fb}(t,\tau) + \xi(t), \tag{5}$$

$$\frac{dN(t)}{dt} = J - \frac{N(t)}{\tau_s} - \frac{G_N(N(t)-N_0)}{1+\varepsilon|E(t)|^2}|E(t)|^2, \tag{6}$$

where $E(t)$ is the complex electric-field amplitude; $N(t)$ is the carrier density; $J$ is the injected current, which is normalized by the lasing threshold current ($J = j \cdot J_{th}$). In the numerical calculations, we add the white Gaussian noise $\xi(t)$ into Eq. (5). We use the laser intensity $I(t) = |E(t)|^2$ as the reservoir response signal. Table I shows the values of the semiconductor laser parameters used in this study. The feedback coefficient $\kappa$ in Eq. (5) is related to the feedback strength $r_3$ by $\kappa = (1 - r_2^2) r_3/(r_2\, \tau_{in})$ [13], where $r_2$ is the reflectivity of the internal laser cavity and $\tau_{in}$ is the optical round-trip time in the internal laser cavity.

TABLE I

Parameter Values Used for Numerical Simulations

| Parameter | Symbol | Value |
|---|---|---|
| Gain coefficient | $G_N$ | $8.04 \times 10^{-13}$ m$^3$s$^{-1}$ |
| Carrier density at transparency | $N_0$ | $1.40 \times 10^{24}$ m$^{-3}$ |
| Injection current at threshold | $J_{th}$ | $9.982 \times 10^{32}$ m$^{-3}s^{-1}$ |
| Normalized injection current | $j$ | 1.1~2.0 |
| Photon lifetime | $\tau_p$ | $1.927 \times 10^{-12}$ s |
| Carrier lifetime | $\tau_s$ | $2.04 \times 10^{-9}$ s |
| Linewidth enhancement factor | $\alpha$ | 3.0 |
| Optical wavelength | $\lambda$ | $1.537 \times 10^{-6}$ m |
| Gain saturation coefficient | $\epsilon$ | $2.0 \times 10^{-23}$ |
| Feedback delay time | $\tau$ | $8.0 \times 10^{-8}$s |
| Sampling frequency of A/D converter | $f_s$ | $2.0 \times 10^{10}$ Hz |
| Cutoff frequency of A/D converter | $f_c$ | $9.9 \times 10^{9}$ Hz |

The modulation signal of the input part is injected into the semiconductor laser by modulating the phase of the feedback signal (*i.e.*, phase modulation scheme) [14]. $E^{fb}(t,\tau)$ represents optical feedback with phase modulation. The phase modulation scheme is expressed as follows,

$$E^{fb}(t,\tau) = E(t-\tau)\exp(-i[\omega\tau + \pi s(t)]), \tag{7}$$

where $s(t)$ is the modulation signal calculated by the input section, $\tau$ is the feedback delay time, and $\omega\tau + \pi s(t)$ is the phase shift due to the delay time and phase modulation. The angular optical frequency of the laser $\omega$ is determined by $\omega = 2\pi c/\lambda$ using the wavelength of the laser $\lambda$.

### *C. Output part*

The output part acquires the node states of the reservoir and is responsible for generating the output signal for reservoir computing. In this scheme, the laser intensity of the semiconductor laser is used to represent the node states (response signals) from the reservoir and is acquired at a constant sampling interval $\theta_s$. The sampled data is low-pass filtered by a third-order Butterworth filter with the cut-off frequency to reduce high-frequency components. The node states are further quantized for approximate reservoir computing (see Sec. III for details).

The weighted linear sum of the reconstructed quantized node states $\hat{x}_k(n)$ is used to optimize the output weights so that the error between the output signal $\hat{y}(n)$ and the target signal $y(n)$ is minimized. The weighted linear sum of the node states is shown below.

$$\hat{y}(n) = \sum_{k=1}^{N} w_k \hat{x}_k(n)\,, \qquad (8)$$

where $w_k$ represents the output weight corresponding to the $k$-th node. Quantization is also applied to the output weights (see Sec. III). The weights are calculated using the least-squares method [5].

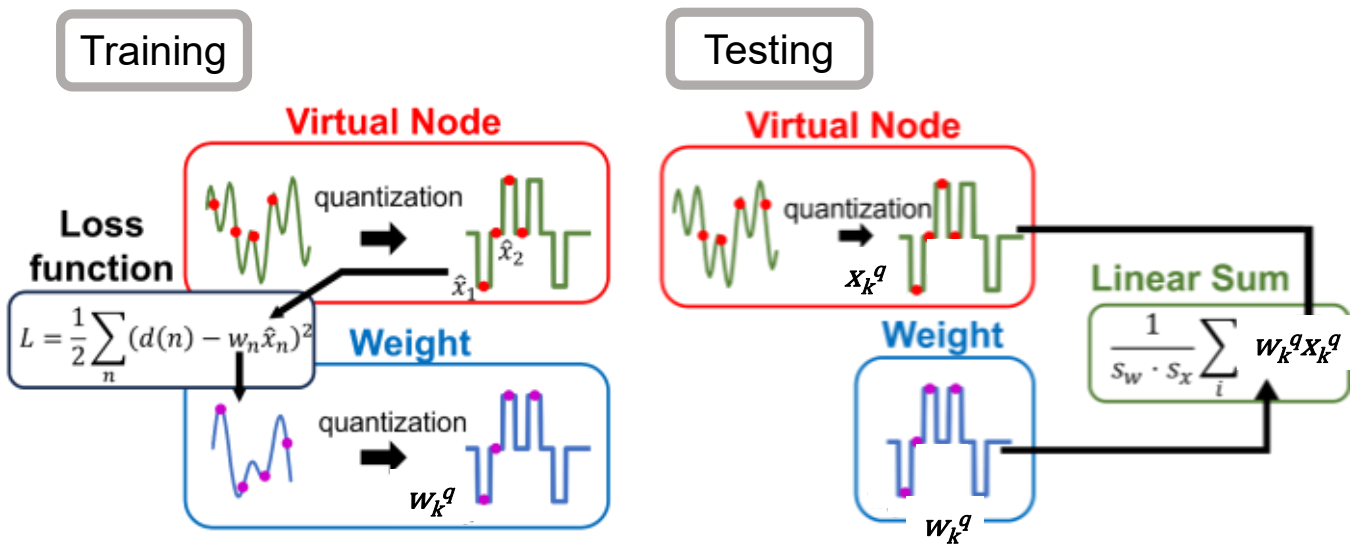

**Fig. 2.** Quantization scheme for training and testing. The node states from the reservoir and the output weights are quantized.

## III. Quantization

Quantization of the node states by an analog-to-digital converter (ADC) is required for photonic reservoir computing. We introduce quantization of the node states in our numerical simulations. We employ a signed symmetric scale quantization scheme [15].

For a $b$-bit signed symmetric quantizer, the real-valued node state $x$ is mapped to the integer code $x^q$ in the range $[-(2^{b-1}-1), 2^{b-1}-1]$, which contains $2^b-1$ symmetric representable integer levels, including zero. For a prescribed real-valued quantization range $[-\alpha_x\sigma_x, \alpha_x\sigma_x]$, where $\sigma_x$ denotes the standard deviation of the node states and $\alpha_x$ is the scaling factor that determines the quantization range, the quantization step is defined as follows.

$$\Delta_x = \frac{\alpha_x \sigma_x}{2^{b-1} - 1}, \qquad (9)$$

The clipping function is defined as

$$clip(x, m, M) = \begin{cases} m\ (x < m) \\ x\ (m \le x \le M) \\ M\ (x > M) \end{cases}, \qquad (10)$$

The node state is first clipped to the prescribed quantization range and then mapped to the integer code according to

$$x^q = round\left[\frac{clip(x, -\alpha_x\sigma_x, \alpha_x\sigma_x)}{\Delta_x}\right], \qquad (11)$$

The corresponding reconstructed real-valued node state $\hat{x}$ is given by

$$\hat{x} = \Delta_x x^q, \qquad (12)$$

Thus, the endpoints $\pm\alpha_x\sigma_x$ are mapped to the maximum-magnitude integer codes $\pm(2^{b-1}-1)$, and zero is represented exactly by $x^q = 0$.

To ensure a consistent quantization model, the reservoir states are quantized using a signed symmetric quantizer with the range $[-\alpha_x\sigma_x, \alpha_x\sigma_x]$ and the quantization step $\Delta_x$. The reservoir state is first clipped to this range and then quantized to the nearest representable level. The appropriate value of $\alpha_x$ is determined from the prediction performance, as discussed in Sec. IV-B.

The weighted linear sum in Eq. (8) can be evaluated using the quantized integer values and their corresponding quantization steps. Similarly, the output weight $w_k$ is quantized to the integer code $w_k^q$, with the corresponding quantization step denoted by $\Delta_w$. Equation (8) can therefore be approximated as

$$\hat{y}(n) = \sum_{k=1}^{N} w_k x_k(n) \approx \Delta_x \Delta_w \sum_{k=1}^{N} w_k^q x_k^q(n), \qquad (13)$$

Here, $x_k^q(n)$ and $w_k^q$ are the quantized integer codes of the node state and output weight at the $k$-th node, respectively, and $\Delta_x$ and $\Delta_w$ are their quantization steps. Equation (13) allows the weighted linear sum to be calculated using integer arithmetic during testing.

To distinguish the quantized integer value from its reconstructed value, $x^q$ denotes the signed integer value of the reservoir state, and the corresponding reconstructed state is obtained using the quantization step $\Delta_x$. The matrix of reservoir states used for ridge-regression training is constructed from the reconstructed states, and the output weights $w$ are calculated in double-precision floating-point arithmetic. After training, the output weights are quantized with a quantization step $\Delta_w$, determined from the maximum absolute value of the trained weights with the symmetric quantization range.

During testing, Eq. (13) uses the quantized integer values $x^q$ and $w^q$, with the factor $\Delta_x\Delta_w$ providing the scaling to the reconstructed numerical output. In the present simulations, the multiply-accumulate operation and accumulation are evaluated in double-precision floating-point arithmetic. Finite-bit quantization is applied only to the reservoir states and output weights.

## IV. Prediction Results

### A. Time-series prediction task

We use the Santa Fe time-series prediction task generated by a far-infrared laser [16], in which specific time-series data are given as input, and the value of one-point-ahead data is predicted in time. We sequentially send the data to the reservoir at the $n$-th point and evaluate the performance by predicting the $(n+1)$-th point. 3000 points are used for training, and 1000 points are used for testing. We use the normalized mean square error (NMSE) to evaluate performance as follows.

$$\text{NMSE} = \frac{1}{L}\sum_{n=1}^{L}\frac{\left(y(n)-\hat{y}(n)\right)^2}{var(y)}, \qquad (14)$$

where $L$ is the number of data points, $y(n)$ is the target signal, and $\hat{y}(n)$ is the predicted signal. We define the success of prediction when NMSE is less than 0.1.

To evaluate the statistical reliability of the prediction results, we perform ten independent calculations for each parameter setting using ten different realizations of the Gaussian noise. The mean NMSE is then calculated by averaging the NMSE values obtained from these ten trials.

### *B. Optimization of quantization range*

We first investigate the appropriate dynamic range for quantization of the node states. Figure 3 shows the prediction error, quantified by the normalized mean square error (NMSE), as a function of the scaling factor $\alpha_x$. The quantization range of the node states is defined as $[-\alpha_x\sigma_x, \alpha_x\sigma_x]$, where $\sigma_x$ denotes the standard deviation of the node states. The minimum NMSE is obtained at $\alpha_x = 1$, corresponding to the quantization range $[-\sigma_x, \sigma_x]$, rather than at a wider range such as $[-3\sigma_x, 3\sigma_x]$. Therefore, we set $\alpha_x = 1$ and use $[-\sigma_x, \sigma_x]$ as the quantization range of the node states in the following calculations.

This choice intentionally clips node-state values outside $[-\sigma_x, \sigma_x]$; however, restricting the quantization range decreases the quantization step size for a fixed number of quantization levels and thereby provides finer resolution within the central region of the node-state distribution. Thus, the selection of the quantization range involves a trade-off between clipping error and quantization resolution. Similar optimization of the quantization range by clipping outliers has been investigated in low-precision neural-network quantization [24], [25], and clipping-based quantization has also been employed in the context of computation-in-memory systems [26].

The results in Fig. 3 indicate that, for the present reservoir-computing prediction task, the benefit of finer quantization resolution within this range outweighs the degradation caused by saturation of relatively large-amplitude node states. We emphasize that the optimum value $\alpha_x = 1$ is specific to the present task and is determined empirically from the results shown in Fig. 3.

### *C. Evaluation of prediction performance for different quantization bits*

Figure 4 shows the results of the chaotic time-series prediction task. The black curve represents the target signal, the red curve represents the predicted signal, and the blue curve corresponds to the prediction error. The numbers of quantization bits for the node states $b_x$ and the output weights $b_w$ are changed simultaneously ($b = b_x = b_w$) to $b = 2$ and 4 bits for Figs. 4(a) and 4(b), respectively. The normalized injection current of the semiconductor laser is fixed at $j = 1.3$, and the quantization range of the node states is set to $[-\sigma_x, \sigma_x]$, corresponding to $\alpha_x = 1$, as determined in Sec. IV-B. For $b = 2$ bits in Fig. 4(a), the NMSE is 2.2, indicating a large prediction error. This can be attributed to the reduced representation capability of the node states due to low-bit quantization. In contrast, for $b = 4$ bits in Fig. 4(b), the NMSE is 0.069, and successful prediction is achieved (NMSE < 0.1). These results demonstrate that the time-series prediction task can be performed with 4-bit quantization of both the node states and output weights.

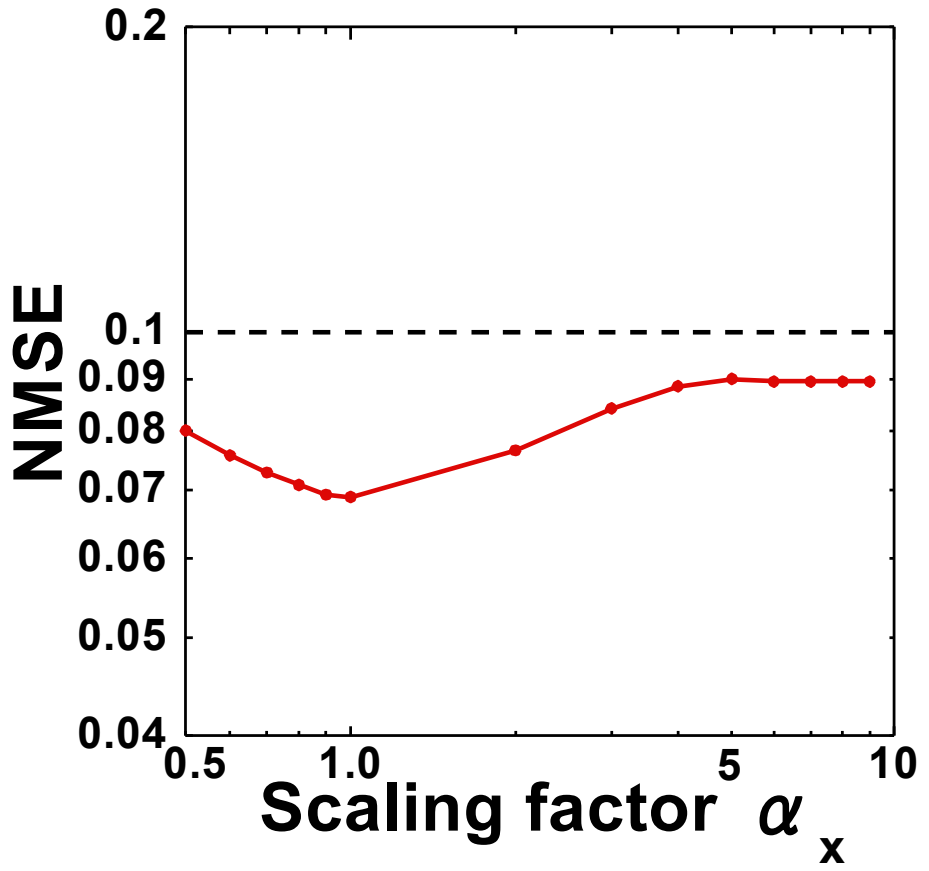

Fig. 3. Normalized mean square error (NMSE) as a function of the scaling factor $\alpha_x$ of the quantization range. The node states are quantized over the range $[-\alpha_x\sigma_x, \alpha_x\sigma_x]$, where $\sigma_x$ denotes the standard deviation of the node states.

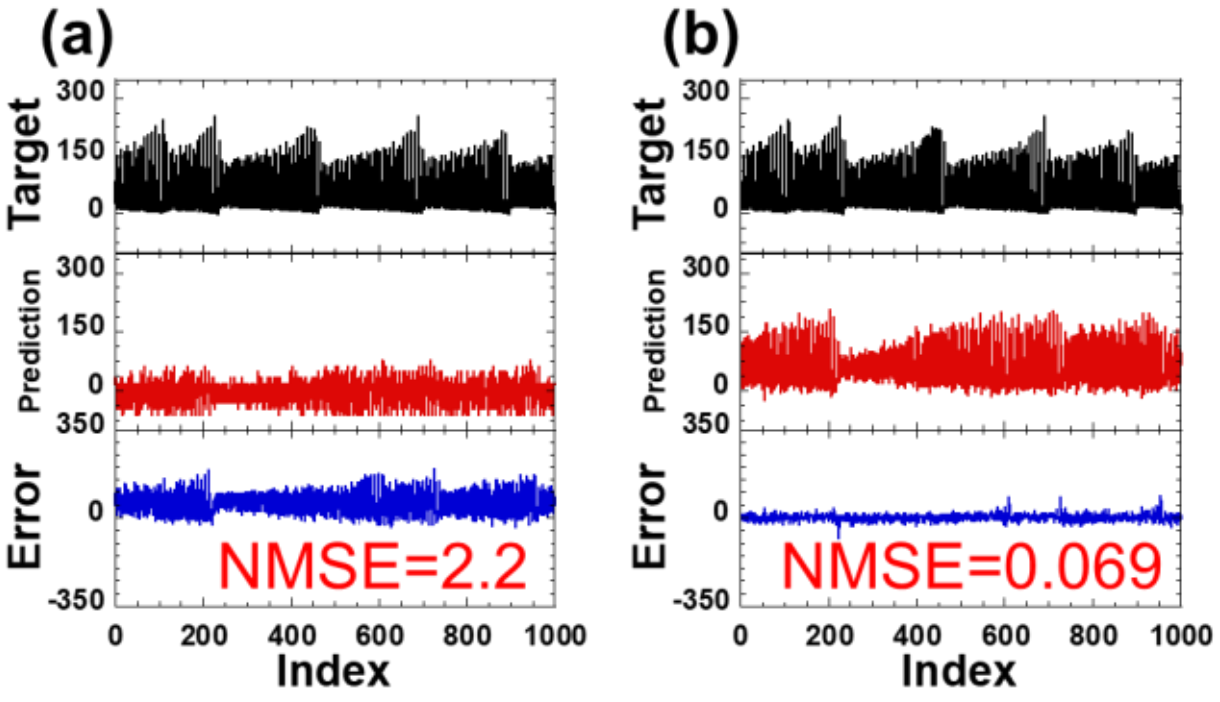

Fig. 4. Results of chaotic time-series prediction task for the number of quantization bits (a) $b = 2$ and (b) $b = 4$. The normalized injection current is fixed at $j = 1.3$.

Figure 5(a) shows the systematic dependence of the prediction error (NMSE) on the number of quantization bits. The blue dotted curve indicates the case when changing the number of quantization bits only for the node states bx, the black solid curve indicates the case when changing the number of quantization bits only for the output weights bw, and the red dashed curve corresponds to the case when changing both bx and bw simultaneously ($b = b_x = b_w$). For the blue dotted curve in Fig. 5(a), the NMSE is less than 0.1 and successful prediction is achieved for bx ≥ 2 bits. This indicates that low-bit quantization of the node states is possible while maintaining high prediction performance. In contrast, for the black solid curve, the NMSE is less than 0.1 for bw ≥ 4 bits, and large NMSEs are obtained for bw = 2 and 3 bits.

(a)

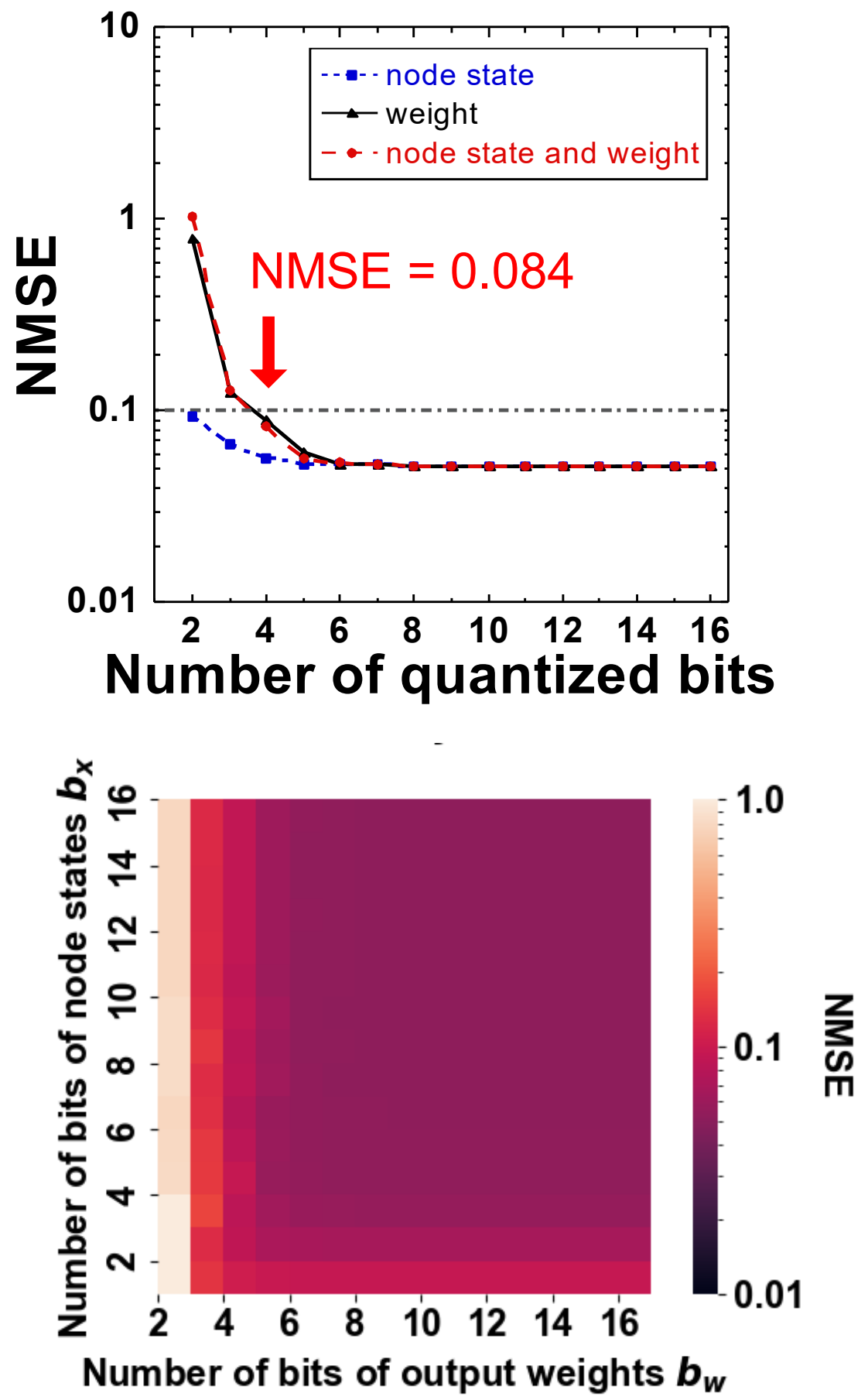

Fig. 5. (a) Prediction error (NMSE) as a function of the number of quantization bits of the node states $b_x$ and the output weights $b_w$. Blue dotted curve: only the node states are quantized. Black solid curve: only the output weights are quantized. Red dashed curve: both the node states and output weights are quantized. (b) Two-dimensional map of NMSE as functions of the number of quantization bits of node states $b_x$ and output weights $b_w$.

This indicates that quantization of the output weights affects the prediction performance more strongly than quantization of the node states. For the red dashed curve, the NMSEs for quantization of both the node states and output weights are similar to those for quantization of only the output weights (black solid curve). From these results, quantization of the output weights influences the prediction performance more strongly, and four- or higher-bit quantization of the output weights is required to maintain prediction performance.

Figure 5(b) shows the two-dimensional map of the NMSE when $b_x$ and $b_w$ are changed. The NMSE is more strongly affected by $b_w$ than by $b_x$. This indicates that the output weights are more crucial in determining the required number of quantization bits.

The node-state precision $b_x$ and output-weight precision $b_w$ are treated as independent parameters because they affect the prediction performance differently. The ADC-related energy is associated with $b_x$, whereas $b_w$ affects the prediction performance of the digital readout.

The stronger degradation caused by output-weight quantization can be attributed to the training procedure. During training, the quantized node states are used to determine the output weights, while the output weights are optimized in 64-bit floating-point precision. Thus, the effect of state quantization is incorporated into the training process and can be partially compensated for by optimizing the output weights. In contrast, output-weight quantization is applied after training and directly perturbs the optimized output weights without subsequent re-optimization. These perturbations directly affect the weighted sum used to generate the output signal and introduce an additional error that cannot be compensated for during training. Consequently, output-weight quantization results in greater degradation of the prediction performance than node-state quantization.

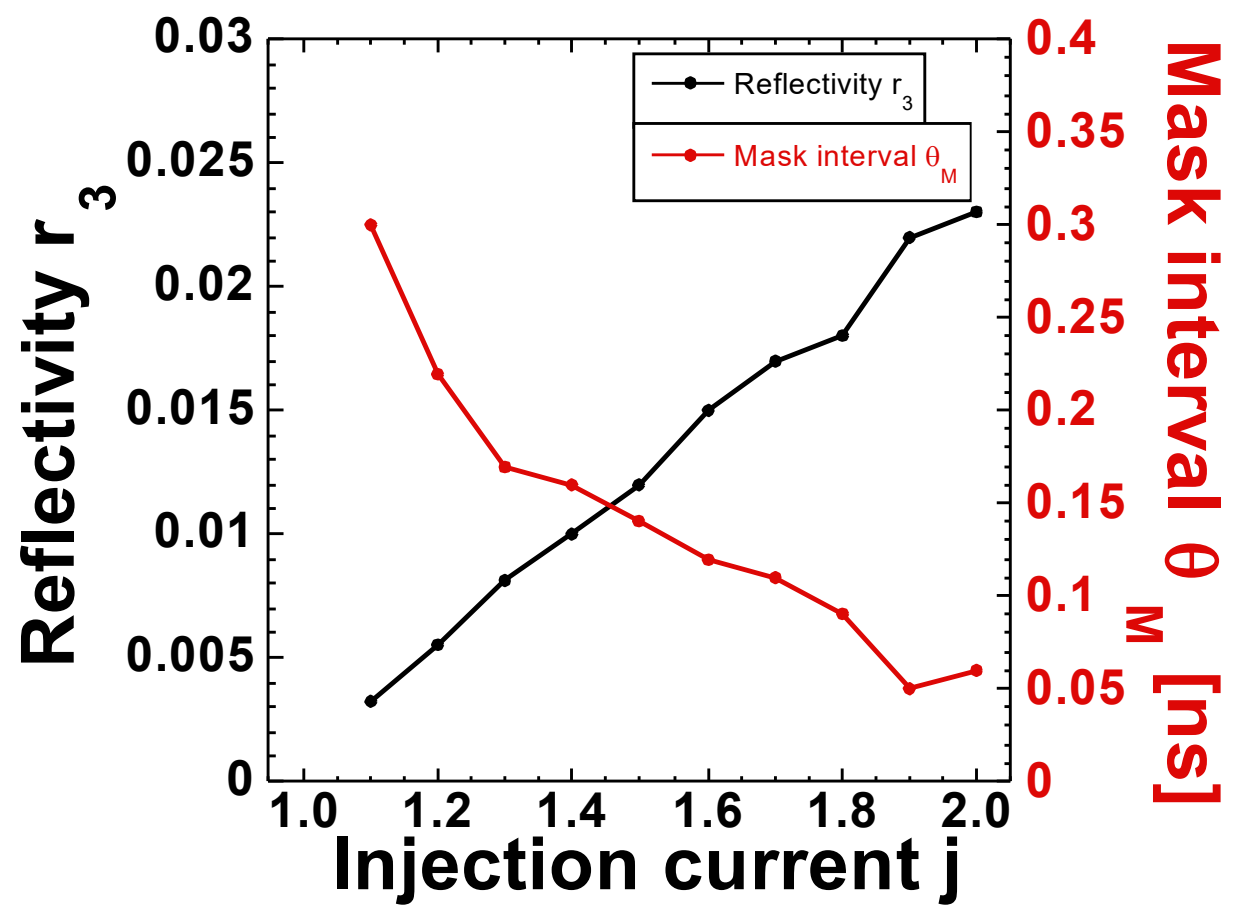

Fig. 6. Optimized feedback strength $r_3$ (black curve) and input-mask interval $\theta_M$ (red curve) as a function of the normalized injection current $j$. for the sampling frequency $f_s$ = 20 GHz.

*D. Evaluation of prediction performance for different injection currents of the semiconductor laser*

We also investigate the effect of the normalized injection current of the semiconductor laser on the prediction performance. For each injection current $j$, the feedback strength $r_3$ and the input-mask interval $\theta_M$ are optimized to achieve a low prediction error, as shown in Fig. 6. The feedback strength $r_3$ determines the dynamical regime of the reservoir and is selected such that the system operates in a stable regime close to the onset of oscillation dynamics. Operation near this regime provides both consistency of the reservoir response and sufficiently complex nonlinear dynamics for temporal information processing. As the injection current $j$ increases, the bifurcation point for the onset of oscillation dynamics shifts, and therefore the optimal value of $r_3$ increases accordingly. In contrast, the input-mask interval $\theta_M$ determines the time scale at which the input mask is applied to the reservoir and affects the temporal response of the reservoir laser. The optimal value of $\theta_M$ decreases with increasing $j$. This behavior originates from the increase in the relaxation oscillation frequency with increasing $j$ [13], which enables the reservoir to respond to faster input modulation. Therefore, a shorter $\theta_M$ becomes suitable at higher injection currents. Based on these considerations, $r_3$ and $\theta_M$ are optimized for each value of $j$, and the NMSE and energy consumption are evaluated using the corresponding optimized parameter values.

Figure 7(a) shows a two-dimensional map of the NMSE when the normalized injection current and the numbers of quantization bits for both the node states and output weights are changed simultaneously ($b = b_x = b_w$). The quantization range of the node states is fixed at [$-\sigma_x$, $\sigma_x$], corresponding to $\alpha_x = 1$, as determined in Sec. IV-B. The sampling frequency is set to $f_s = 20$ GHz. The dark region indicates better prediction performance. For a large number of quantization bits, the NMSE converges to a low value, and good prediction performance is obtained. In addition, better prediction performance is achieved at larger normalized injection currents. However, only limited improvement is obtained for 2-bit and 3-bit quantization even when the normalized injection current is increased. Figure 7(b) shows the same plot when the sampling frequency is reduced from 20 to 2 GHz. Very similar results to those in Fig. 7(a) are obtained.

From these results, successful prediction (NMSE < 0.1) is achieved for the conditions indicated by the blue square in Figs. 7(a) and 7(b), namely, a normalized injection current of $j \geq 1.3$ and a quantization bit number of $b \geq 4$. For example, the prediction performance is obtained with an NMSE of 0.0931 ± 0.0096 at $f_s = 2$ GHz, $j = 1.3$, and $b = 5$ ($= b_x = b_w$), where the value represents the mean ± standard deviation over ten trials with different Gaussian noise realizations.

*E. Evaluation of prediction performance for different sampling frequencies*

To investigate the influence of the sampling frequency $f_s$ on the prediction performance, we evaluate the NMSE for different sampling frequencies ranging from 1 to 20 GHz, as shown in Fig. 8. To account for the bandwidth limitation associated with the sampling process, the original temporal waveforms are low-pass filtered with a cutoff frequency of ($f_s/2$ - 0.1) GHz before sampling, which mimics the frequency limitation of an analog-to-digital converter operating at the corresponding sampling frequency. For each sampling frequency $f_s$ and injection current $j$, the feedback strength $r_3$ and the input-mask interval $\theta_M$ are optimized to achieve a low prediction error.

Figure 8 shows the NMSE as a function of the injection current $j$ for different values of $f_s$. Although decreasing $f_s$ reduces the temporal resolution and the number of available sampling points within the fixed feedback delay time $\tau = 80$ ns, sufficiently low prediction errors can still be obtained over the investigated range of $f_s$. In particular, the NMSE becomes lower than 0.1 for $j \geq 1.3$ at all sampling frequencies from 1 to 20 GHz. At $f_s = 2$ GHz, the NMSE also remains below 0.1 for $j \geq 1.3$. These results demonstrate that reducing the sampling frequency from 20 to 2 GHz does not significantly degrade the prediction performance when an appropriate injection current and optimized reservoir parameters are used.

(a)

(b)

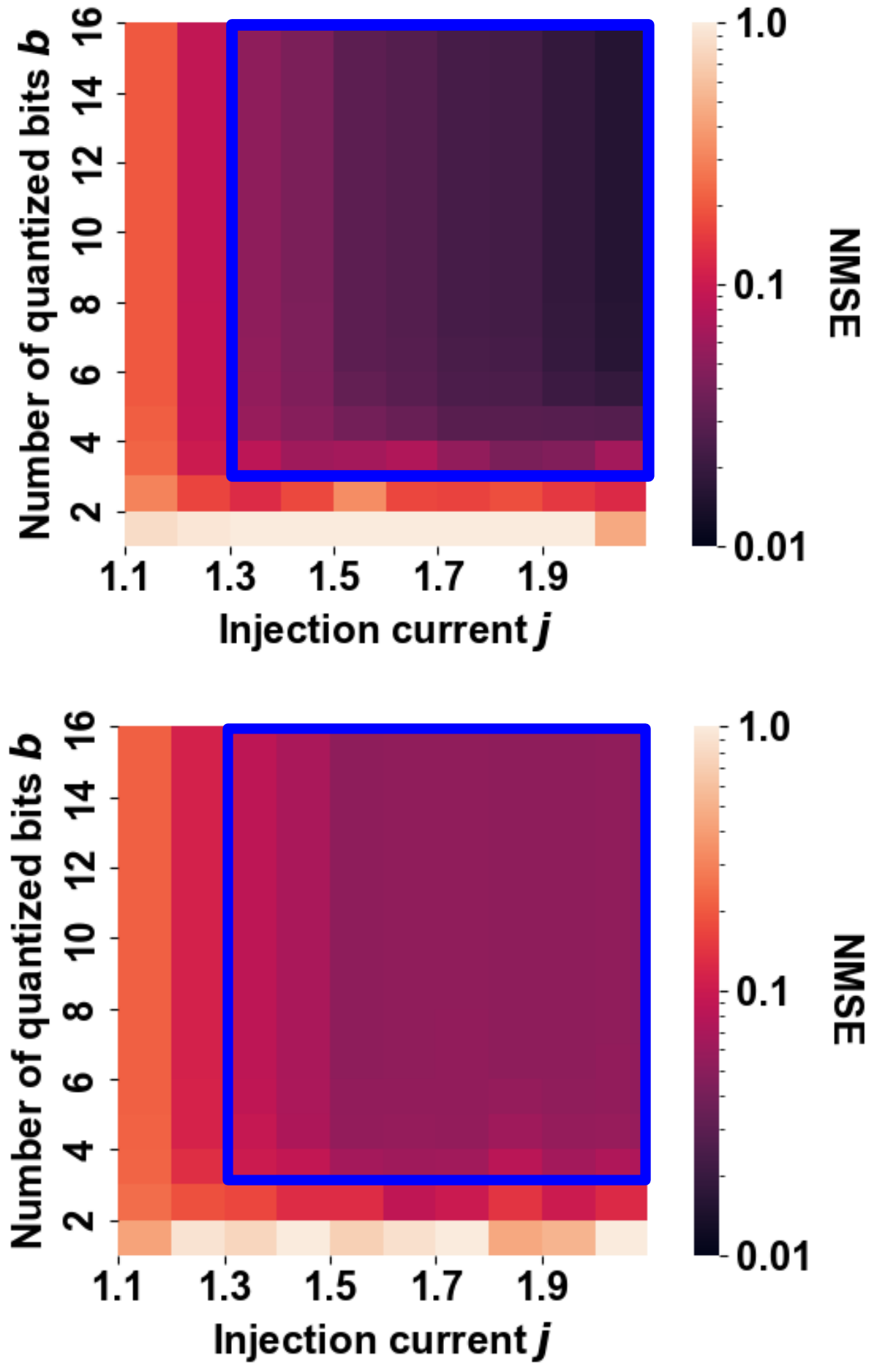

Fig. 7. Two-dimensional map of NMSE as functions of the number of quantization bits of both node states and output weights ($b = b_x = b_w$) and the injection current of the semiconductor laser $j$. The sampling frequency is (a) $f_s = 20$ GHz and (b) $f_s = 2$ GHz. The region within the blue square corresponds to the successful prediction (NMSE < 0.1).

The number of nodes $N$ is determined from the feedback delay time $\tau$, the optimized input-mask interval $\theta_M$, and the sampling interval $\theta_s$. For a given sampling frequency $f_s$, the sampling interval is defined as $\theta_s = 1/f_s$, while the maximum number of nodes is limited to $N_{max} = 200$. When $\theta_M > \theta_s$, multiple sampling points are available within each mask interval, and one of these sampling points is selected as the node state for each mask interval. In this case, the number of nodes is determined by $N = \tau/\theta_M$. In contrast, when $\theta_M \leq \theta_s$, the temporal resolution is limited by the sampling interval, and all available sampled states within one feedback-delay period $\tau$ are used as node states. The number of nodes is then determined by $N = \tau/\theta_s$. If the calculated number of nodes exceeds $N_{max}$, the additional node states are discarded so that $N$ does not exceed 200 within $\tau$. Using $\tau$ = 80 ns and the optimized values of $\theta_M$, the resulting numbers of nodes are $N$ = 200 for $f_s$ = 5, 10, and 20 GHz, $N$ = 160 for $f_s$ = 2 GHz, and $N$ = 80 for $f_s$ = 1 GHz. Thus, even though reducing $f_s$ decreases the number of accessible node states, particularly at 1 and 2 GHz, the results in Fig. 8 confirm that an NMSE below 0.1 can still be achieved for $j \geq 1.3$.

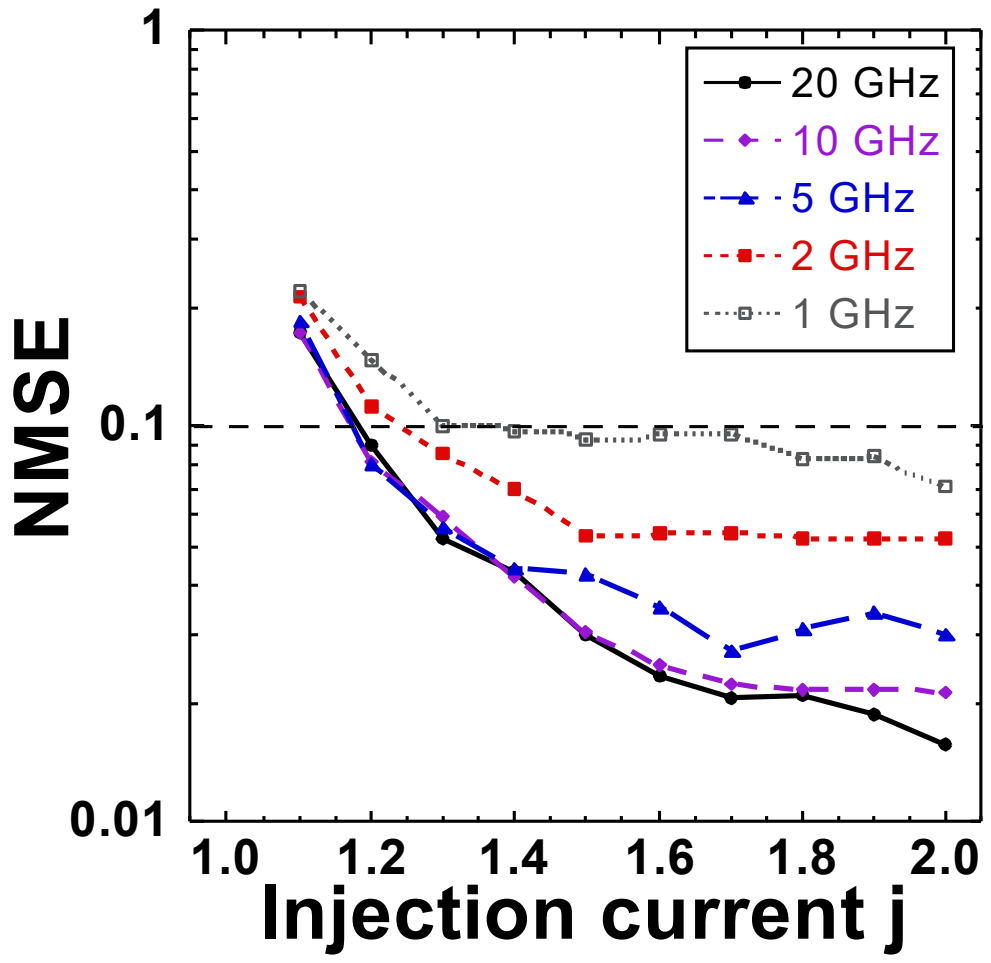

Fig. 8. Prediction error (NMSE) as a function of the injection current $j$ for sampling frequencies $f_s$ ranging from 1 to 20 GHz. Black solid curve with circles: $f_s$ = 20 GHz, purple dotted-dashed curve with diamonds: $f_s$ = 10 GHz, blue dashed curve with triangles: $f_s$ = 5 GHz, red dotted curve with squares: $f_s$ = 2 GHz, and gray dotted curve with white squares: $f_s$ = 1 GHz.

## V. Evaluation of Energy Consumption

### A. Walden Model for Analog-to-Digital Converter

In this section, we estimate the energy consumption of the Analog-to-Digital Converter (ADC) used for acquiring the reservoir response signal using the Walden figure-of-merit model [17,18]. This model relates the ADC energy per conversion to the intrinsic converter resolution and is expressed by the following equation.

$$\frac{P}{f_s} = \mathrm{FOM_w} \times 2^{\mathrm{ENOB}} \tag{15}$$

Here, $\mathrm{FOM_w}$ is the Walden figure of merit [18], $P$ is the ADC power consumption and $f_s$ is the ADC sampling frequency. $P/f_s$ represents the energy per ADC conversion. ENOB is the effective number of bits of the ADC required to digitize the reservoir output [19]. ENOB is estimated as follows.

$$\sigma_q^2 = \frac{1}{12}\left(\frac{\alpha_x \sigma_x}{2^{b-1} - 1}\right)^2 \tag{16}$$

$$\mathrm{SNDR} = 10\log_{10}\left(\frac{\sigma_x^2}{\sigma_q^2 + \sigma_N^2}\right) \tag{17}$$

$$\mathrm{ENOB} = \frac{\mathrm{SNDR} - 1.76}{6.02} \tag{18}$$

where $\sigma_x^2$ is the variance of the node states of the reservoir, $\sigma_N^2$ is the variance of the node-state noise arising from intrinsic semiconductor-laser noise, $\sigma_q^2$ is the quantized noise estimated using the quantization-noise approximation, $b$ is the number of quantization bits, and SNDR stands for Signal-to-Noise and Distortion Ratio of the reservoir node states. SNDR is affected by the number of quantization bits of the node states.

The intrinsic ENOB of an ADC is distinguished from the SNDR of the analog reservoir output, which includes intrinsic semiconductor-laser noise. In this study, the reservoir-signal SNDR is used as an approximate indicator of the effective resolution required for A/D conversion, rather than as a direct measure of the intrinsic ADC ENOB. This approach assumes that the ADC resolution is selected according to the reservoir-signal quality to avoid unnecessarily high conversion precision beyond the intrinsic noise level of the laser output.

### B. Evaluation of Energy Consumption of Analog-to-Digital Converter

Figure 9(a) shows the NMSE and the energy consumption of the ADC per sample (denoted as $P/f_s$) when the number of quantization bits for both the node states and the output weights is changed simultaneously ($b = b_x = b_w$). For $b$ = 2 bits, NMSE is large and the energy consumption is low. As $b$ is increased, NMSE decreases and the energy consumption increases, then both values saturate at large $b$. Although $b_x = b_w$ is used in this figure, only $b_x$ is associated with the ADC energy, whereas $b_w$ affects only the prediction performance in the present energy analysis.

Figure 9(b) shows the NMSE and the estimated ADC energy per conversion $P/f_s$ as the normalized injection current $j$ is varied at $f_s$ = 20 GHz. As $j$ increases, the reservoir-signal SNDR increases, indicating an improvement in signal quality relative to the intrinsic laser noise. Under the adopted SNDR-based energy estimation model, this improvement corresponds to a higher equivalent ADC resolution and, consequently, higher estimated ADC energy consumption. This dependence reflects the modeling assumption relating the reservoir-signal SNDR to the ADC resolution requirements, rather than a direct influence of $j$ on the intrinsic ADC characteristics. The criterion for successful prediction (NMSE < 0.1) is satisfied for $j \geq 1.3$.

(a)

(b)

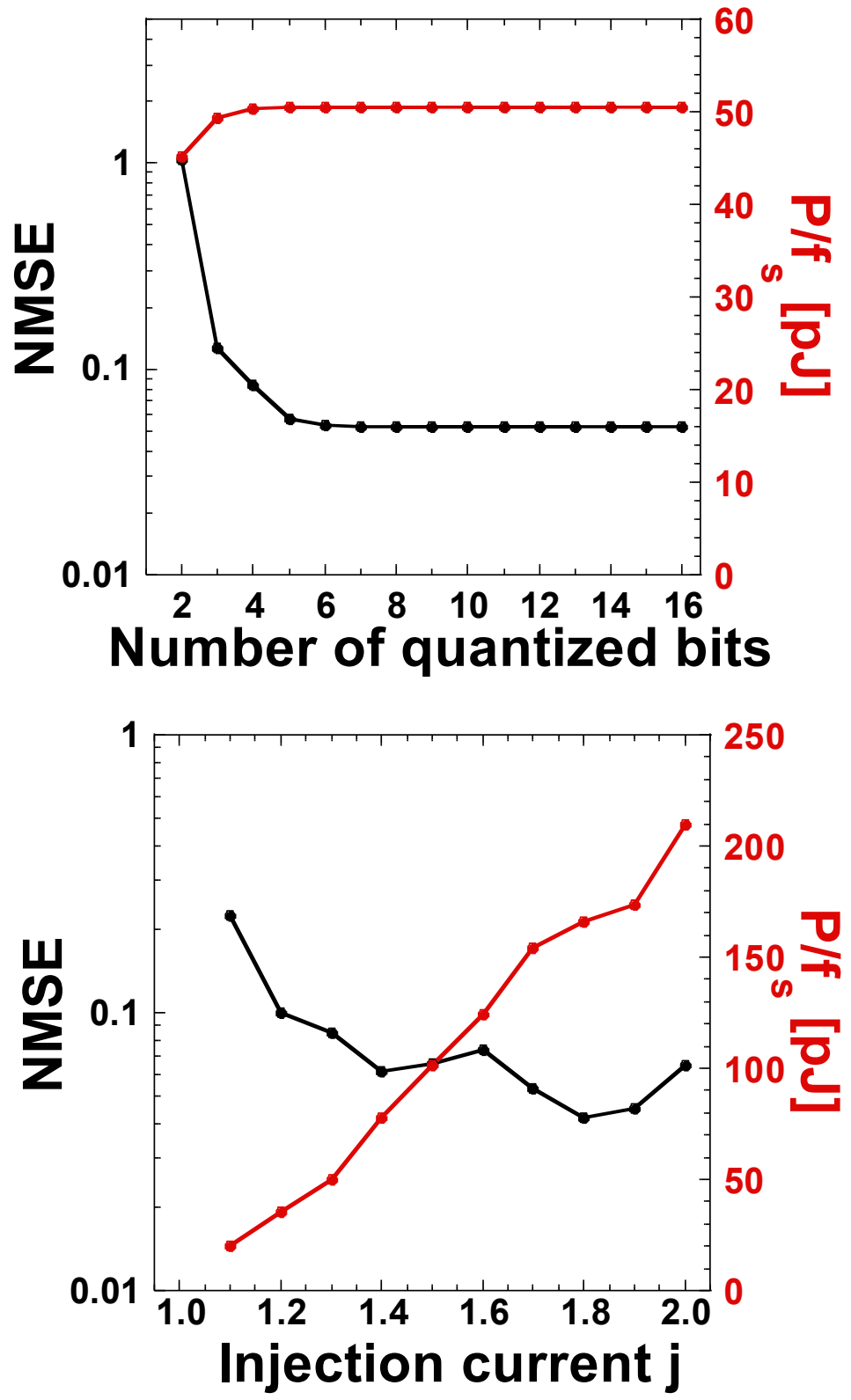

Fig. 9. Prediction error (NMSE, black curve) and energy consumption per sample of the ADC ($P/f_s$, red curve) as a function of (a) the number of quantization bits for both the node states and the output weights $b$ (= $b_x$ = $b_w$) and (b) the normalized injection current of the semiconductor laser $j$. The sampling frequency is fixed at $f_s$ = 20 GHz.

The normalized injection current $j$ can indirectly influence the ADC resolution requirements at the system level. As $j$ increases, the relative contribution of intrinsic semiconductor-laser noise to the reservoir signal can decrease, making quantization noise relatively more important. Consequently, a higher ADC resolution may be beneficial for preserving the improved reservoir-signal quality. Under the adopted modeling assumption, the SNDR-derived equivalent bit resolution is used in the Walden model to estimate the ADC energy consumption. This estimate is model-dependent and should not be interpreted as the actual energy consumption of a specific physical ADC. Therefore, the dependence of the estimated ADC energy on $j$ reflects the assumed relationship between reservoir-signal quality and ADC resolution, rather than a direct dependence of the intrinsic ADC characteristics on $j$.

Figure 10(a) shows the two-dimensional diagram of the energy consumption per sample of the ADC ($P/f_s$) when changing the number of quantization bits of both node states and output weights $b$ (= $b_x$ = $b_w$) and the normalized injection current of the semiconductor laser $j$. The horizontal axis corresponds to $j$, and the vertical axis corresponds to $b$. The color indicates $P/f_s$. The sampling frequency is fixed at $f_s$ = 20 GHz. For Fig. 10(a), higher energy consumption is required for higher $j$. In contrast, the energy consumption does not change significantly with $b$. This suggests that low normalized injection current effectively results in low energy consumption of the ADC.

(a)

(b)

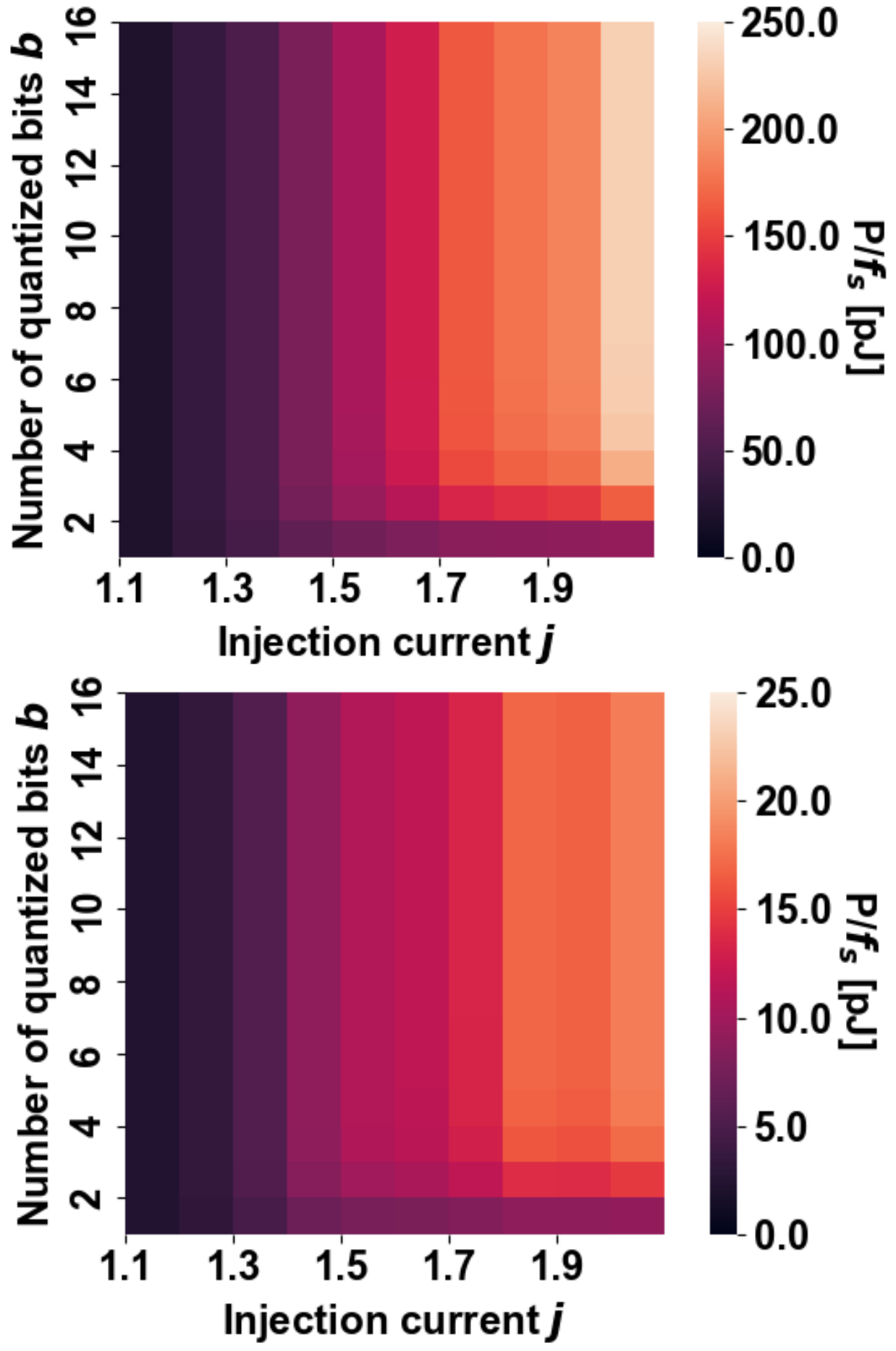

Fig. 10. Two-dimensional map of energy consumption per sample of the ADC ($P/f_s$) as functions of the number of quantization bits of both node states and output weights $b$ (= $b_x$ = $b_w$) and the normalized injection current of the semiconductor laser $j$. The sampling frequency is fixed at (a) $f_s$ = 20 GHz and (b) $f_s$ = 2 GHz.

Figure 10(b) shows the same figure when the sampling frequency is reduced from 20 to 2 GHz. Very similar results are obtained, except for the energy consumption value, which is reduced by one order of magnitude in Fig. 10(b). Among the parameter conditions of successful prediction (NMSE < 0.1, see Fig. 7(b)), the mean ADC energy consumption is 5.332 ± 0.0147 pJ at $f_s$ = 2 GHz, $j$ = 1.3, and $b$ = 5 (= $b_x$ = $b_w$).

To clarify the individual contributions of the approximate-computing parameters to the energy reduction, we estimate the energy consumption of the ADC by independently varying $f_s$, $j$, and $b$, as shown in Fig. 11. Figure 11(a) shows the dependence on the quantization precision $b$ at different sampling frequencies $f_s$. The energy consumption increases with $b$ at low bit precision and tends to saturate for $b \geq 4$ bits, indicating that $b$ has a limited impact on the ADC energy consumption beyond 4 bits. Figure 11(b) shows the dependence on the injection current $j$ at different $f_s$. The energy consumption increases approximately linearly with $j$, indicating that $j$ has a significant impact on the ADC energy consumption. In addition, the energy consumption increases with $f_s$ in both Figs. 11(a) and 11(b). Under the adopted SNDR-based ADC energy estimation model, reducing $j$ and $f_s$ provides the major contributions to the estimated ADC energy reduction, whereas reducing $b$ provides a comparatively limited benefit for $b \geq 4$ bits.

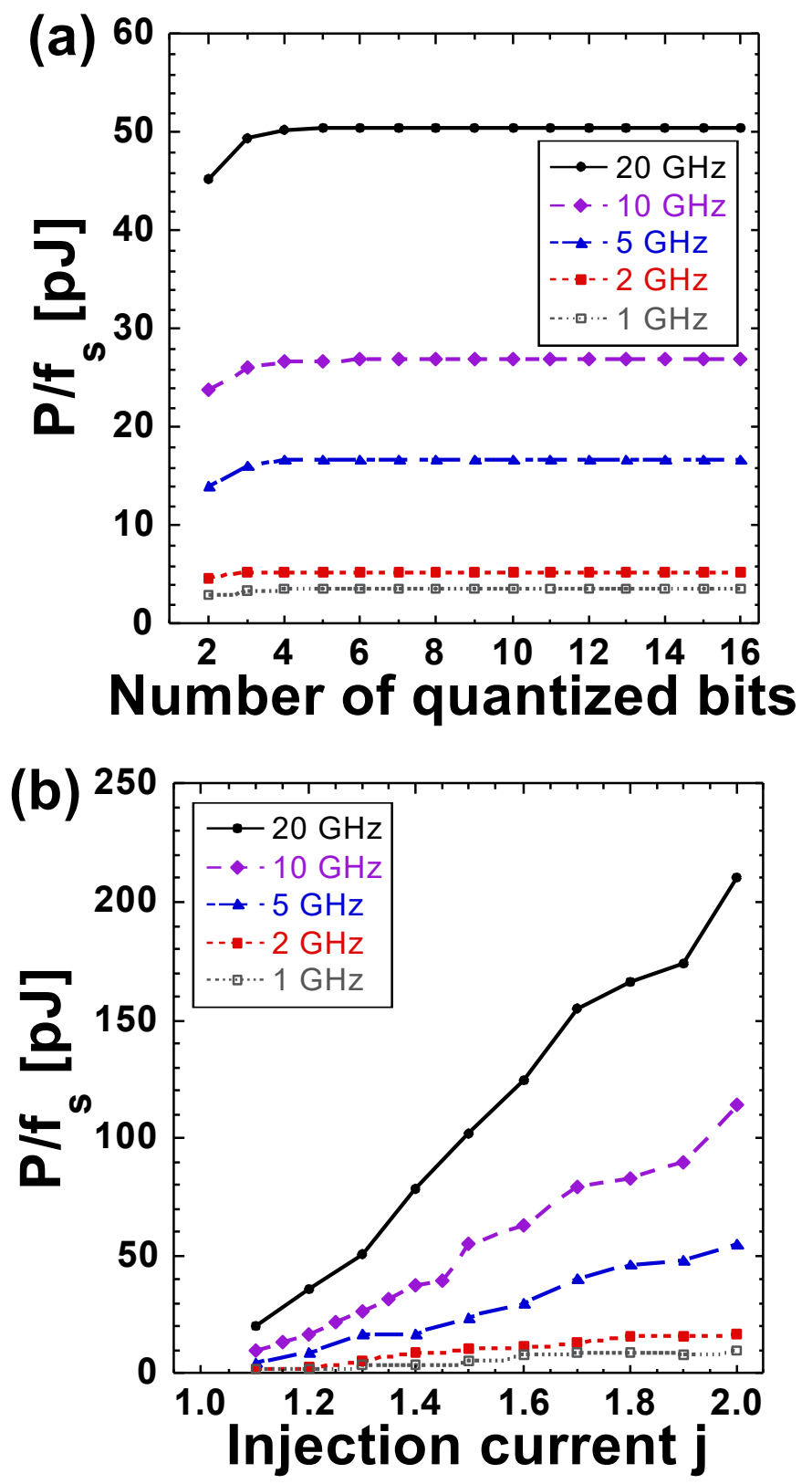

Fig. 11. (a) Energy consumption of the ADC as a function of the quantization precision $b$ at different sampling frequencies $f_s$. (b) Energy consumption of the ADC as a function of the injection current $j$ at different $f_s$. Black solid curve with circles: $f_s$ = 20 GHz, purple dotted-dashed curve with diamonds: $f_s$ = 10 GHz, blue dashed curve with triangles: $f_s$ = 5 GHz, red dotted curve with squares: $f_s$ = 2 GHz, and gray dotted curve with white squares: $f_s$ = 1 GHz.

In the present SNDR and ADC energy estimates, Eq. (16) indicates the conventional quantization-noise approximation, using the quantization step defined in Eq. (9). This approximation assumes a uniformly distributed quantization error within one quantization step and does not include clipping error from node-state values outside the prescribed range. Clipping is applied in the quantized reservoir-computing simulations, so its effect is reflected in the reported prediction NMSE; however, its contribution is not included in the quantization-noise term used to estimate SNDR and ADC energy. Omitting this contribution could overestimate the reservoir-signal SNDR relative to an estimate that includes both granular and clipping errors. A quantitative evaluation of the total quantization error, including clipping, and its effect on the SNDR-based ADC energy estimates is left for future work.

### *B. Evaluation of Total Energy Consumption*

Next, we evaluate energy consumption of other experimental components. Figure 12 shows the experimental setup of photonic reservoir computing with a semiconductor laser [14]. We consider the fiber-based experimental apparatus to estimate energy consumption per sample as approximate reservoir computing. The light emitted from the semiconductor laser (LD) propagates in a feedback loop through a circulator, a fiber coupler, an attenuator, and a phase modulator (MOD). The input signal for the reservoir is generated by a Digital-to-Analog Converter (DAC) in an arbitrary waveform generator and fed into the feedback loop via an electrical amplifier (Amp) and MOD. The response signal of the reservoir laser is converted to electrical signals by a photodetector (PD), and sent to an ADC in a digital oscilloscope for the measurement of the node states of the reservoir.

We estimate the energy consumption of ADC, DAC, MOD, Amp, PD, and LD using the reference information [18-23], summarized in Table II. For MOD, Amp, PD, and LD, energy consumption is calculated and these values are divided by the sampling frequency $f_s$ to evaluate the energy consumption per sample $P/f_s$.

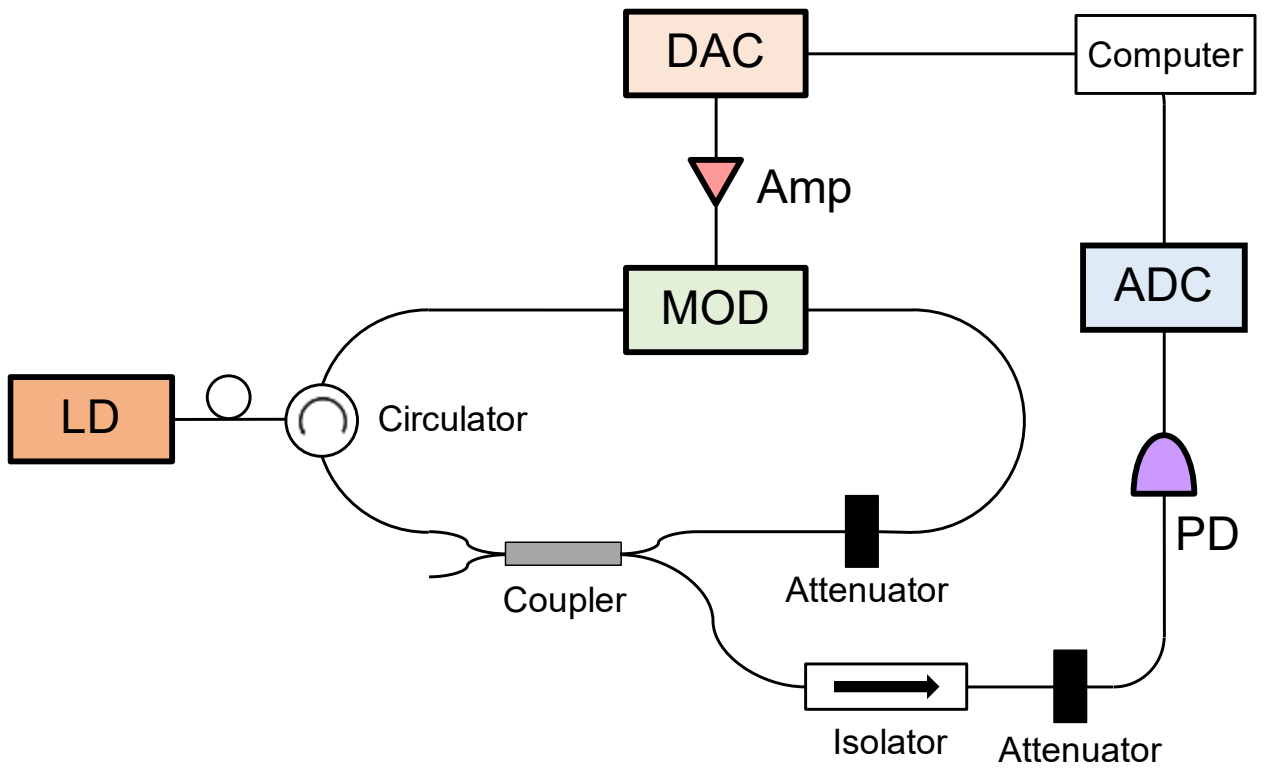

Fig. 12. Schematic of experimental setup for approximate reservoir computing with a semiconductor laser with feedback light [14]. The energy consumption per sample of the following experimental apparatus is evaluated. ADC: Analog-to-Digital Converter, DAC: Digital-to-Analog Converter, MOD: phase modulator, Amp: electric amplifier, PD: photodetector, LD: semiconductor laser.

Figure 13(a) shows the energy consumption per sample $P/f_s$ of the six components and the total energy consumption as the normalized injection current $j$ is varied at $f_s$ = 20 GHz. The variation in the estimated ADC energy with $j$ arises from the adopted SNDR-based energy estimation model, in which the equivalent ADC resolution is determined from the reservoir-signal SNDR. Therefore, this variation should be interpreted as a model-dependent estimate associated with the reservoir-signal quality, rather than a direct dependence of the intrinsic ADC energy consumption on $j$. The successful-prediction region (NMSE < 0.1) is $j \geq 1.3$.

TABLE II
EXPERIMENTAL APPARATUS EVALUATED FOR ENERGY CONSUMPTION

| Experimental apparatus | Symbol | References |
|---|---|---|
| Analog-to-Digital converter | ADC | [18] |
| Digital-to-Analog converter | DAC | [20] |
| Modulator | MOD | [21] |
| Amplifier | Amp | [22] |
| Photodetector | PD | [22] |
| Semiconductor Laser | LD | [23] |

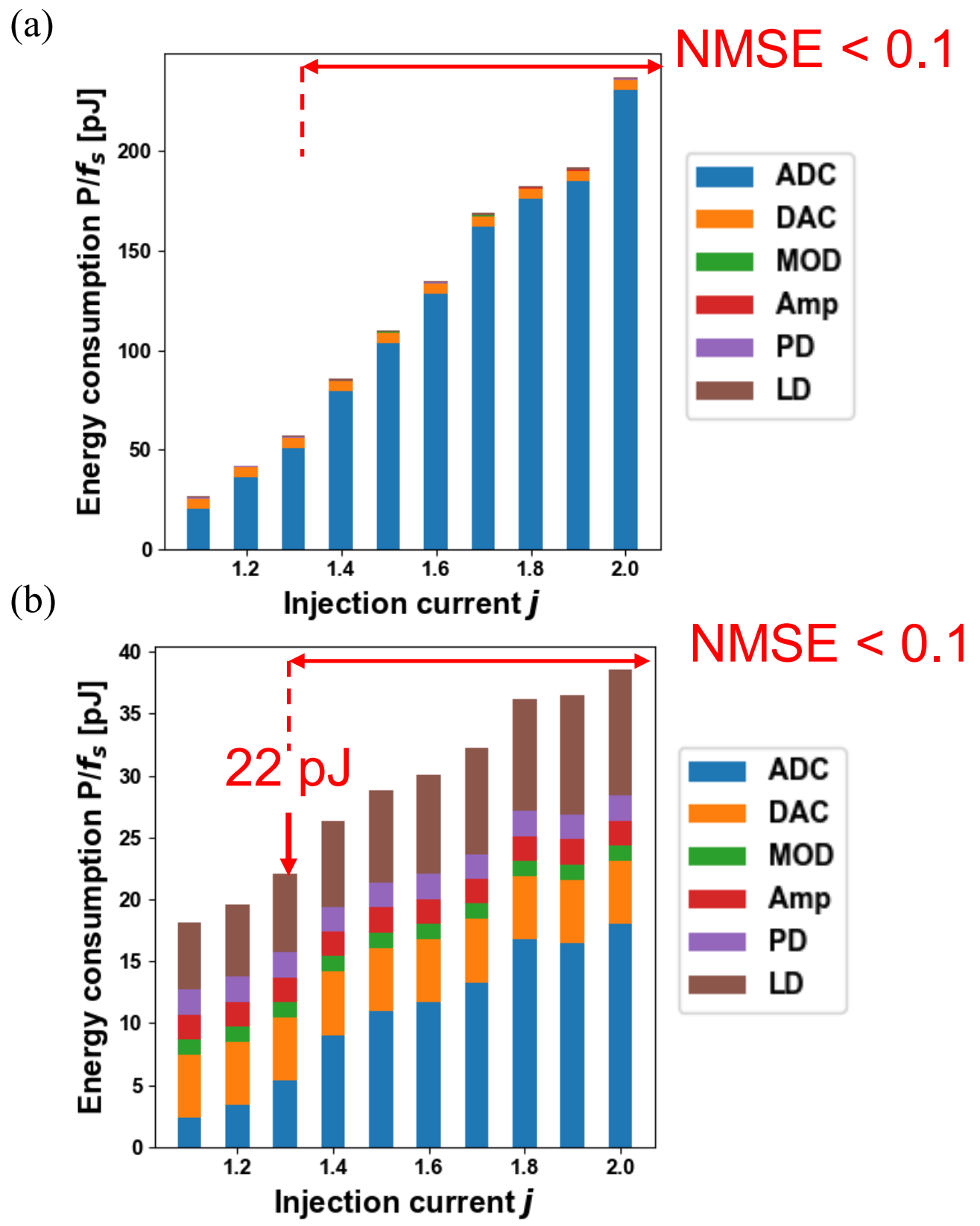

Fig. 13. Energy consumption per sample $P/f_s$ as a function of the normalized injection current of the semiconductor laser $j$ for the sampling frequency of (a) $f_s$ = 20 GHz and (b) $f_s$ = 2 GHz. The number of quantization bits for both the node states and the output weights is fixed at $b$ = 5 bits (= $b_x$ = $b_w$). Blue: ADC (analog-to-digital converter), orange: DAC (digital-to-analog converter), green: MOD (phase modulator), red: Amp (electric amplifier), purple: PD (photodetector), and brown: LD (semiconductor laser).

Figure 13(b) shows $P/f_s$ for the six experimental components as a function of $j$ when the sampling frequency is reduced to $f_s$ = 2 GHz. At low $f_s$, several experimental components become dominant, such as LD (brown in Fig. 13(b)) and DAC (orange), rather than ADC (blue). The energy consumption per sample of LD increases for a lower $f_s$ because the energy consumption is estimated from the constant energy divided by $f_s$. On the contrary, the energy consumption per sample of ADC decreases due to a lower $FOM_w$ for a lower $f_s$.

In addition, the total energy consumption can be reduced as $j$ is reduced. To satisfy the condition of successful prediction (NMSE < 0.1), the injection current needs to be set to $j \geq 1.3$, as shown in the red arrow in Fig. 13(b). In this condition of successful prediction, the minimum total energy consumption per sample is 22 pJ at $j$ = 1.3.

Figure 14 shows the comparison of the total energy consumption per sample $P/f_s$ for different sampling frequencies $f_s$. The injection current is set to $j$ = 1.3 and the number of quantization bits is set to $b$ = 5 bits (= $b_x$ = $b_w$). For Fig. 14, the relative contributions to energy consumption among the six experimental components change for different $f_s$, i.e., the energy consumption of ADC is dominant for a higher $f_s$, whereas the energy consumption of LD is dominant for a lower $f_s$. We found the minimum total energy consumption of 22 pJ at $f_s$ = 2 GHz in Fig. 14. Therefore, $f_s$ can be optimized for reducing the total energy consumption per sample.

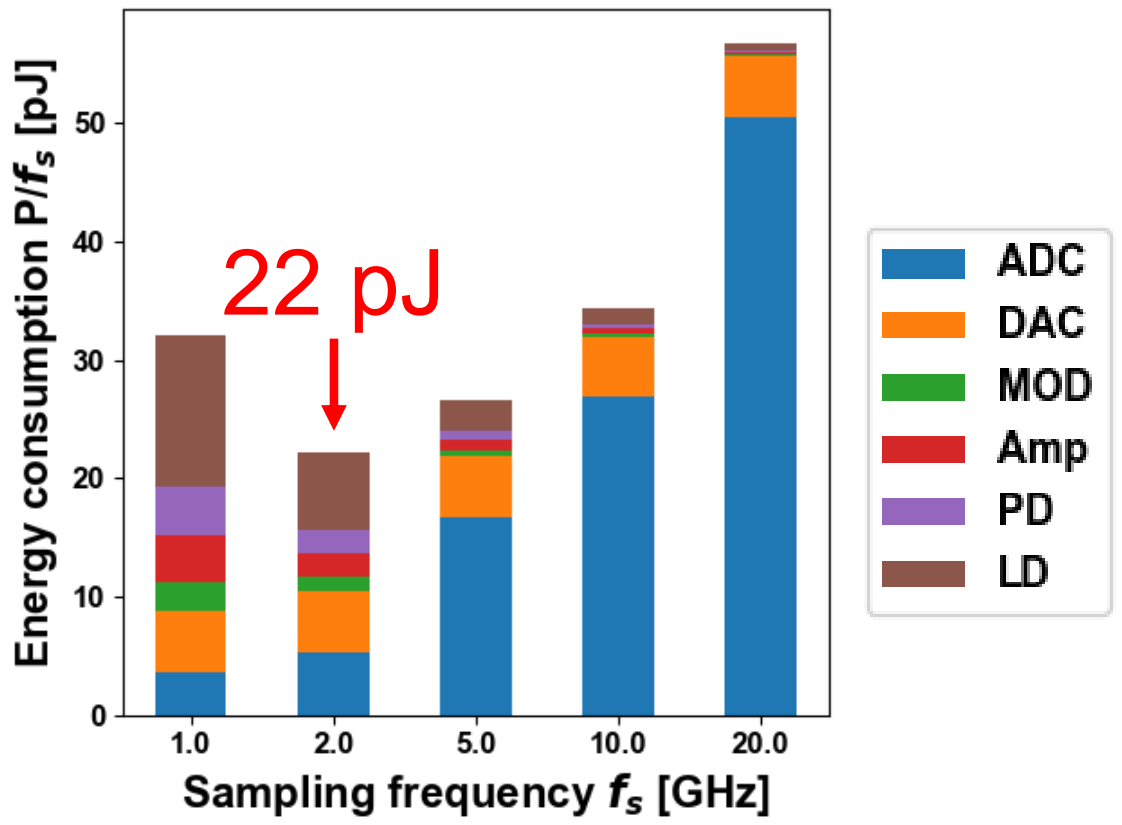

Fig. 14. Energy consumption per sample $P/f_s$ as a function of the sampling frequency $f_s$. The injection current is set to $j$ = 1.3. The number of quantization bits for both the node states and the output weights is fixed at $b$ = 5 bits (= $b_x$ = $b_w$). Blue: ADC (analog-to-digital converter), orange: DAC (digital-to-analog converter), green: MOD (phase modulator), red: Amp (electric amplifier), purple: PD (photodetector), and brown: LD (semiconductor laser).

## VI. Discussion of Energy Consumption

The energy consumption is estimated to be 22 pJ per sampled node. At a sampling frequency of $f_s$ = 2 GHz and a delay time of 80 ns, each input datum is represented by $N$ = 160 sampled nodes. The energy consumption per input datum is estimated to be 3.52 nJ (22 pJ × 160). Thus, the energy per sampled node should be distinguished from the energy required to process one input datum. The energy per sampled node is used here as a hardware-level metric, analogous to the energy-per-sample metric reported for photonic computing systems [27], whereas the energy per input datum represents the energy associated with all 160 sampled nodes corresponding to one input datum.

It should be noted that the number of physical ADC conversions can differ from the number of retained virtual nodes. Assuming continuous ADC operation during the delay interval of τ = 80 ns (100% duty cycle), the number of physical ADC conversions is given by $N_{conv} = f_s\tau$, corresponding to 80, 160, 400, 800, and 1600 conversions at $f_s$ = 1, 2, 5, 10, and 20 GHz, respectively, whereas the numbers of retained virtual nodes are 80, 160, 200, 200, and 200. The reported ADC energy represents the estimated energy per physical conversion, whereas the ADC energy per input datum includes all conversions performed during the delay interval, including those whose samples are subsequently discarded. However, at $f_s$ = 2 GHz, all physical ADC samples are retained as virtual nodes, and their acquisition energy is fully included in the energy calculation. Thus, $f_s$ = 2 GHz provides the lowest estimated total energy per sample among the investigated sampling frequencies.

TABLE III
COMPARISON OF PARAMETER VALUES AND TOTAL ENERGY CONSUMPTION

| Parameter | Symbol | Proposed | Baseline [14] |
|---|---|---|---|
| Sampling frequency [GHz] | $f_s$ | 2 | 20 |
| Normalize injection current | $j$ | 1.3 | 2.0 |
| Quantization precision [bit] | $b_x$, $b_w$ | $b_x$ = 5, $b_w$ = 5 | $b_x$ = 8, $b_w$ = 8 |
| Total energy consumption per sample [pJ] | $P/f_s$ | 22 | 237 |

For comparison, we define a Ref. [14]-inspired baseline configuration using the operating parameters listed in Table III. The energy consumption of this baseline is estimated using the same component-level energy model and energy-accounting boundary as those used for the proposed configuration. Therefore, the baseline energy of 237 pJ/sample is a model-estimated value obtained in the present study and is not an experimentally measured or directly reported value in Ref. [14]. The sampling frequency, normalized injection current, and quantization precision assumed for this baseline are explicitly listed in Table III. Under the component-level energy model adopted in this study, the estimated total energy per sample decreases from 237 pJ/sample for the baseline to 22 pJ/sample for the proposed operating condition, corresponding to a 91% reduction in energy per sample. The comparison quantifies the reduction in energy per sample estimated within the common energy-accounting framework adopted in the present study.

We evaluate the performance metrics and optimal operating condition as follows. At the optimal operating condition of $f_s$ = 2 GHz with $N$ = 160, the sampling interval is $\theta s = 1/f_s$ = 0.5 ns, resulting in a reservoir processing time per input datum of $T = N\theta s$ = 80 ns. Following the operation-counting convention of Ref. [10], the equivalent computational workload for a one-dimensional input and a single reservoir channel is $11N$ multiply-accumulate (MAC) operations per input datum. Therefore, the equivalent throughput is $11N/T$ = (11×160)/(80 ns) = 22 GMAC/s. With an energy consumption of 3.52 nJ per input datum, the corresponding average power during reservoir processing is 3.52 nJ/80 ns = 0.044 W, yielding an energy efficiency of 22 GMAC/s/0.044 W = 500 GMAC/J. Equivalently, the energy per equivalent operation is 3.52 nJ/(11×160) = 2.0 pJ/MAC. These performance metrics and the corresponding operating parameters are summarized in Table IV.

TABLE IV
SUMMARY OF OPERATING PARAMETERS AND PERFORMANCE METRICS AT THE OPTIMAL OPERATING CONDITION

| Parameter or performance metric | Value |
|---|---|
| Sampling frequency, $f_s$ | 2.0 GHz |
| Sampling interval, $\theta_s$ | 0.50 ns |
| Feedback delay time, $\tau$ | 80.0 ns |
| Normalized injection current, $j$ | 1.3 |
| Number of nodes, $N$ | 160 |
| Reservoir processing time per input datum, $T$ | 80.0 ns |
| Node-state precision, $b_x$ | 5 bits |
| Output-weight precision, $b_w$ | 5 bits |
| Prediction error (NMSE) | 0.0931 |
| Equivalent throughput | 22 GMAC/s |
| Energy per sample | 22 pJ/sample |
| Energy per input datum | 3.52 nJ/input datum |
| Energy efficiency | 500 GMAC/J |
| Energy per equivalent operation | 2.0 pJ/MAC |

The equivalent computational workload is evaluated as $11N$ multiply-accumulate (MAC) operations per input datum, where $N$ is the number of nodes, following the operation-counting convention of Ref. [10].

For comparison, Nakajima et al. [10] reported a system-level energy efficiency of 15.9–33.2 GMAC/J, whereas Sunada et al. [11] estimated a potential energy consumption of 0.015 fJ/MAC for passive multimode-waveguide processing. For reference, the corresponding values estimated for the present system are 500 GMAC/J and 2.0 pJ/MAC, respectively. However, these numerical values should not be interpreted as a direct ranking of the energy efficiency of the different architectures because the energy-accounting boundaries differ substantially among the studies. The estimate in Ref. [10] includes the power consumption of the entire experimental system, including optical and electronic components, whereas the potential value in Ref. [11] is based on the optical input power required for passive photonic processing and excludes the peripheral electronic equipment that dominated the power consumption of the experimental implementation. Similarly, our estimate includes the energy consumption of the components considered in the present model (ADC, DAC, modulator, amplifier, photodetector, and laser) but does not represent the total wall-plug energy consumption of a complete experimental system. Therefore, the values reported in Refs. [10] and [11] are presented as reference benchmarks rather than as directly comparable measures of energy efficiency.

## VII. Conclusion

We introduced a concept of approximate reservoir computing and investigated the performance of chaotic time-series prediction task. We changed the number of quantization bits of the node states and the output weights and investigated the prediction performance. We also evaluated the energy consumption of the experimental apparatus used for approximate reservoir computing. The ADC-related energy is associated with b_x, whereas b_w affects the prediction performance of the digital readout and is outside the present energy-accounting boundary. Under the component-level energy model adopted in this study, the optimized operating condition yields an estimated energy consumption of 22 pJ/sample, corresponding to a model-estimated reduction of approximately 91% relative to the Ref. [14]-inspired baseline of 237 pJ/sample evaluated using the same energy-accounting framework. This comparison is model based and should not be interpreted as an experimentally demonstrated or strictly performance-matched energy reduction relative to Ref. [14]. We also evaluated the equivalent throughput of 22 GMAC/s, energy efficiency of 500 GMAC/J, and the energy per equivalent operation of 2.0 pJ/MAC in the proposed reservoir-computing system. The reported energy values were estimated numerically using a component-level model rather than measured directly from a physically implemented system.

Our findings are promising for the real-world experimental implementation of approximate reservoir computing to reduce energy consumption in edge-computing applications.

## References

[1] K. Kitayama, M. Notomi, M. Naruse, K. Inoue, S. Kawakami, and A. Uchida, "Novel frontier of photonics for data processing - Photonic accelerator," *APL Photonics*, vol. 4, pp. 090901 (2019).

[2] Y. Shen, N. C. Harris, S. Skirlo, M. Prabhu, T. Baehr-Jones, M. Hochberg, X. Sun, S. Zhao, H. Larochelle, D. Englund, and M. Soljacic, "Deep learning with coherent nanophotonic circuits," *Nature Photonics*, vol. 11, pp. 441-446 (2017).

[3] H. Jaeger and H. Haas, "Harnessing nonlinearity: Predicting chaotic systems and saving energy in wireless communication," *Science*, vol. 304. no. 5667, pp. 78-80 (2004).

[4] G. Tanaka, T. Yamane, J. B. Héroux, R. Nakane, N. Kanazawa, S. Takeda, H. Numata, D. Nakano, and A. Hirose, "Recent advances in physical reservoir computing: A review," *Neural Networks*, vol. 115, pp. 100–123 (2019).

[5] L. Appeltant, M. C. Soriano, G. Van der Sande, J. Danckaert, S. Massar, J. Dambre, B. Schrauwen, C. R. Mirasso, and I. Fischer, "Information processing using a single dynamical node as complex system," *Nature Communications*, vol. 2, Art. no. 468 (2011).

[6] D. Brunner, M. C. Soriano, C. R. Mirasso, and I. Fischer, "Parallel photonic information processing at gigabyte per second data rates using transient states," *Nature Communications*, vol. 4, Art. no. 1364 (2013).

[7] S. Mittal, "A survey of techniques for approximate computing," *Association for Computing Machinery*, vol. 48, no. 4, pp. 1-33 (2016).

[8] B. Moons, B. D. Brabandere, L. V. Gool and M. Verhelst, "Energy-efficient ConvNets through approximate computing," *2016 IEEE Winter Conference on Applications of Computer Vision (WACV)*, pp.1-8 (2016).

[9] M. C. Soriano, S. Ortin, D. Brunner, L. Larger, C. R. Mirasso, I. Fischer, and L. Pesquera, "Optoelectronic reservoir computing: tackling noise-induced performance degradation," *Optics Express*, vol. 21, no. 1, pp. 12–20 (2013).

[10] M. Nakajima, K. Tanaka, and T. Hashimoto, "Scalable reservoir computing on coherent linear photonic processor," *Communications Physics*, vol. 4, Art. no. 20 (2021).

[11] S. Sunada and A. Uchida, "Photonic neural field on a silicon chip: large-scale, high-speed neuro-inspired computing and sensing," *Optica*, vol. 8, no. 11, pp. 1388-1396 (2021).

[12] R. Lang and K. Kobayashi, "External optical feedback effects on semiconductor injection laser properties," *IEEE Journal of Quantum Electronics*, vol. 16, no. 3, pp. 347-355 (1980).

[13] A. Uchida, *Optical Communication with Chaotic Lasers, Applications of Nonlinear Dynamics and Synchronization*. Wiley-VCH, Weinheim (2012). ISBN: 978-3-527-40869-6

[14] K. Kanno, A. Haya and A. Uchida, "Reservoir computing based on an external-cavity semiconductor laser with optical feedback modulation," *Optics Express*, vol. 30, no. 19, pp. 34218-34238 (2022).

[15] H. Wu, P. Judd, X. Zhang, M. Isaev and P. Micikevicius, "Integer quantization for deep learning inference: principles and empirical evaluation," *arXiv preprint* arXiv:2004.09602 (2020).

[16] A. S. Weigend and N. A. Gershenfeld, "Results of the time series prediction competition at the Santa Fe Institute," *IEEE International Conference on Neural Networks*, vol. 3, pp. 1786-1793 (1993).

[17] R. H. Walden, "Analog-to-digital converter survey and analysis," *IEEE Journal on Selected Areas in Communications*, vol. 17, no. 4, pp. 539-550 (1999).

[18] B. Murmann, "ADC Performance Survey 1997-2023," [Online]. Available: https://github.com/bmurmann/ADC-survey.

[19] W. Kester, "Taking the mystery out of the infamous formula, "SNR= 6.02 n+ 1.76 dB," and why you should care," *Analog Devices*, MT-001 (2005).

[20] B. Sedighi, M. Khafaji, and J. C. Scheytt. "8-bit 5 GS/s D/A converter for multi-gigabit wireless transceivers," *2011 6th European Microwave Integrated Circuit Conference*, pp. 192–195. IEEE (2011).

[21] Tektronix, "Manual for Arbitrary Waveform Generators AWG70001A and AWG70002A," (2015). https://www.tek.com/ja/manual/awg70001a-awg70002a

[22] X. Zheng, D. Patil, J. Lexau, F. Liu, G. Li, H. Thacker, Y. Luo, I. Shubin, J. Li, J. Yao, P. Dong, D. Feng, M. Asghari, T. Pinguet, A. Mekis, P. Amberg, M. Dayringer, J. Gainsley, H. F. Moghadam, E. Alon, K. Raj, R. Ho, J. E. Cunningham, and A. V. Krishnamoorthy, "Ultra-efficient 10 Gb/s hybrid integrated silicon photonic transmitter and receiver," *Optics Express*, vol. 19, no. 6, pp. 5172–5186 (2011).

[23] DFB Laser Diode, NTT Electronics, KELD1C5GAAA, https://www.ntt-innovative-devices.com/en/products/photonics/pdf/NLK1E5GAAA.pdf

[24] R. Banner, Y. Nahshan, and D. Soudry, "Post training 4-bit quantization of convolutional networks for rapid-deployment," Advances in Neural Information Processing Systems, vol. 32, 2019.

[25] R. Zhao, Y. Hu, J. Dotzel, C. De Sa, and Z. Zhang, "Improving neural network quantization without retraining using outlier channel splitting," Proceedings of the 36th International Conference on Machine Learning, Proceedings of Machine Learning Research, vol. 97, pp. 7543-7552, 2019.

[26] N. Misawa, R. Yamaguchi, A. Yamada, T. Wang, C. Matsui, and K. Takeuchi, "Design methodology of compact edge vision transformer CiM considering non-volatile memory bit precision and memory error tolerance," Japanese Journal of Applied Physics, vol. 63, pp. 03SP05 (2024).

[27] X. Meng, D. Kong, K. Kim, Q. Li, P. Dong, I. J. Cox, C. Lioma, and H. Hu, "Digital-analog hybrid matrix multiplication processor for optical neural networks," Nature Communications, vol. 16, Art. no. 7465 (2025).

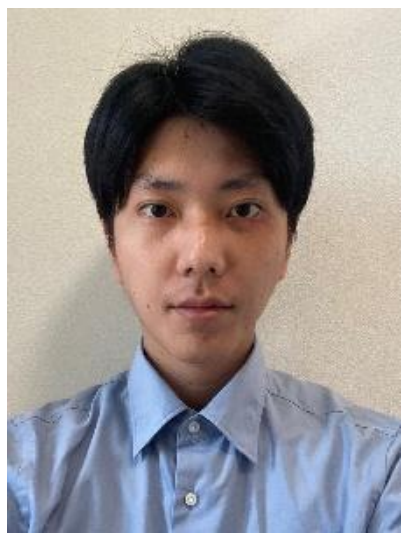

**Tatsuki Ito** received the B.E. and M.E. degrees in Information and Computer Sciences from Saitama University, Saitama, Japan, in 2023 and 2025, respectively.

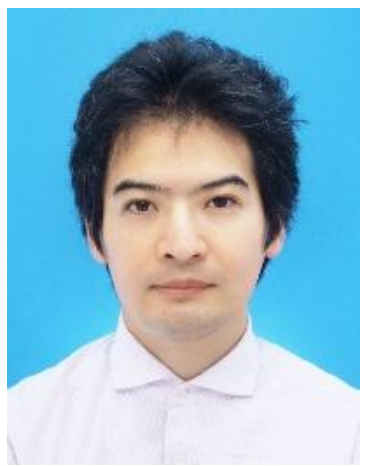

**Kazutaka Kanno** received the B.E., M.E., and Ph.D. degrees in Information and Computer Sciences from Saitama University, Saitama, Japan, in 2009, 2011, and 2014, respectively. In 2014, he joined the Department of Electronics Engineering and Computer Science, Fukuoka University, Fukuoka, Japan, as an Assistant Professor. In 2018, he joined the Department of Information and Computer Sciences, Saitama University, Saitama, Japan, as an Assistant Professor, where he has been an Associate Professor since 2023.

His research interests include nonlinear laser dynamics and its applications in photonic computing.

Dr. Kanno is a member of the Institute of Electronics, Information and Communication Engineers of Japan, the Japan Society of Applied Physics, and the Laser Society of Japan.

**Satoshi Kawakami** (Member, IEEE) received the B.S., M.S., and Ph.D. degrees in computer science from Kyushu University, Fukuoka, Japan, in 2012, 2014, and 2019, respectively. From 2014 to 2016, he was a Hardware Engineer with Bosch Corporation, Japan. From 2016 to 2019, he was an Academic Researcher with Kyushu University, Fukuoka, Japan. From 2019 to 2022, he was an Assistant Professor with Kyushu University. Since 2022, he has been an Associate Professor with the Faculty of Information Science and Electrical Engineering, Kyushu University, Fukuoka, Japan.

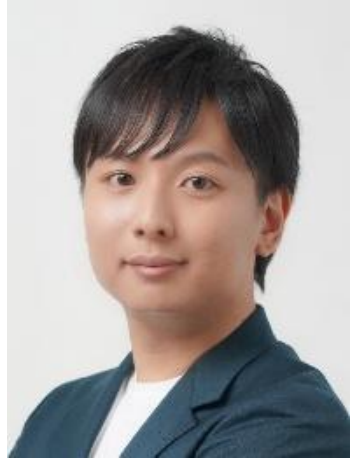

He is currently working on computer architecture for emerging computing devices, with particular emphasis on photonic computing systems and superconducting computing circuits. His research explores how the physical properties of emerging devices, including optical propagation, wavelength multiplexing, analog noise, and cryogenic operation, can be exploited as computational resources rather than regarded merely as implementation constraints. His current interests include photonic accelerators for neural networks, optical computing architectures for recursive and iterative processing, hardware-aware training and fine-tuning for analog accelerators, compute-in-wire and compute-on-arrival architectures, and hardware/software co-design methodologies for post-Moore computing systems.

Dr. Kawakami is a member of IPSJ, IEICE, and ACM.

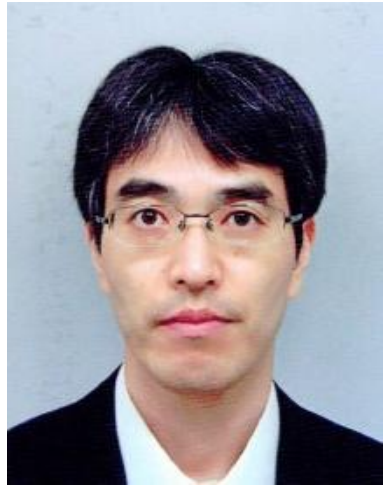

**Atsushi Uchida** (Senior Member, IEEE) received the B.S., M.S., and Ph.D. degrees in electrical engineering from Keio University, Yokohama, Japan, in 1995, 1997, and 2000, respectively. In 2000, he joined the Department of Electronics and Computer Systems, Takushoku University, Tokyo, Japan, where he was a Research Associate. He was a Visiting Researcher with the ATR Adaptive Communications Research Laboratories, Kyoto, Japan, from 2001 to 2002. He was a JSPS Postdoctoral Fellow with the University of Maryland, College Park, USA, from 2002 to 2004. He was a Lecturer with the Department of Electronics and Computer Systems, Takushoku University, Tokyo, Japan, from 2005 to 2008. He was an Associate Professor with the Department of Information and Computer Sciences, Saitama University, Saitama, Japan, from 2008 to 2015. Since 2015, he has been a Professor with the Department of Information and Computer Sciences, Saitama University, Saitama, Japan.

He is currently working on synchronization of chaotic lasers and their applications for optical secure communications, secure key generation using chaotic lasers for cryptography, fast physical random number generation using chaotic lasers, reservoir computing using complex photonics, photonic decision making, and synchronization of chaos and consistency in nonlinear dynamical systems.

Dr. Uchida is a Senior Member of Optica and the Laser Society of Japan. He is a member of the American Physical Society, the Institute of Electronics, Information, and Communication Engineers of Japan, the Japan Society of Applied Physics, and the Optical Society of Japan.